\ifx\fiverm\undefined
        \newfont\fiverm{cmr5}
\fi
\input prepictex
\input pictex
\input postpictex
\catcode`\@=11
\@input{picmore.tex}

\documentstyle[12pt]{article}

\def\labelmark{}

\def\void{}
\newenvironment{formula}[1]{\def\labelname{#1}
\ifx\void\labelname\def\junk{\begin{displaymath}}
\else\def\junk{\begin{equation}\label{\labelname}}\fi\junk}%
{\ifx\void\labelname\def\junk{\end{displaymath}}
\else\def\junk{\end{equation}}\fi\junk\labelmark\def\labelname{}}
{\ifx\void\labelname\def\junk{\end{array}\end{displaymath}}
\else\def\junk{\end{array}\right.\end{equation}}
\fi\junk\labelmark\def\labelname{}\def\junk{}
}
\newenvironment{formulae}[1]{\def\labelname{#1}
\ifx\void\labelname\def\junk{\begin{displaymath}}
\else\def\junk{\begin{eqnarray}\label{\labelname}}\fi\junk}%
{\ifx\void\labelname\def\junk{\end{displaymath}}
\else\def\junk{\end{eqnarray}}\fi\junk\labelmark\def\labelname{}}

\newcommand{\beq}{\begin{formula}}
\newcommand{\eeq}{\end{formula}}
\newcommand{\beqa}{\begin{formulae}}
\newcommand{\eeqa}{\end{formulae}}
\newcommand{\eq}[1]{(\ref{#1})}

\newcommand{\ra}{\rightarrow}

\newcommand{\eps}{\varepsilon}
\newcommand{\NP}[1]{ {\it Nucl.~Phys.} {\bf #1}}
 
\newcommand{\PL}[1]{ {\it Phys.~Lett.} {\bf #1}}

\newcommand{\PR}[1]{ {\it Phys.~Rev.} {\bf #1}}

\newcommand{\PTP}[1]{ {\it Prog.~Theor.~Phys.} {\bf #1}}

\newcommand{\MPL}[1]{ {\it Mod.~Phys.~Lett.} {\bf #1}}
\newcommand{\IJMP}[1]{ {\it Int.~J.~Mod.~Phys.} {\bf #1}}
\newcommand{\JP}[1]{ {\it J.~Phys.} {\bf #1}:\  Math.~Gen.~}
\newcommand{\ZP}[1]{ {\it Z.~Phys.} {\bf #1}}

\newcommand{\JMP}[1]{ {\it J. Math.~Phys.} {\bf #1}}
\newcommand{\WS}[1]{ (World Scientific Publishing, #1)}

\begin{document}
\begin{titlepage}
\setcounter{page}{1}
\renewcommand{\thefootnote}{\fnsymbol{footnote}}

%\rightline{ }
\begin{flushright}
KEK Preprint 2000-5\\
hep-th/0003268
%\mbox{\phantom{NBI-HE-97-34}}\\
%\mbox{\phantom{hep-th/9709047}}\\
%manuscript
\end{flushright}

\vspace{13mm}
\begin{center}
{\Large FFZ realization of the deformed super Virasoro algebra\\
--- Chaichian-Pre\v{s}najder type} 
\vspace{17mm}

{\large Ahmed Jellal 
\footnote{E-mail: oudrhiri@fsr.ac.ma }}
\,\sc{and}\,
{\large Haru-Tada Sato   
\footnote{E-mail: haru@post.kek.jp
%\\ Address after April, 2000: 
}}\\
\vspace{5mm}
{\em $^{}$ Laboratory of Theoretical Physics, Faculty of Science\\
Ibn Battouta Street, P.O. Box 1014, Rabat-Morocco} \\
\vspace{5mm}
{\em $^{}$ Institute of Particle and Nuclear Studies \\
High Energy Accelerator Research Organization (KEK), Tanashi Branch \\
Midori-cho 3-2-1, Tokyo 188-8501, Japan}
\end{center}

\vspace{5mm}

\begin{abstract}
The $q$-deformed super Virasoro algebra proposed by Chaichian and 
Pre\v{s}najder is examined. Presented is the realizations by the FFZ 
algebra (the magnetic translation algebra) defined on a two-dimensional 
lattice with a supersymmetric Hamiltonian. 
\end{abstract}

\vspace{8mm}
\vfill
\begin{flushleft}
\end{flushleft}

\end{titlepage}
\newpage
%%%%%%%%%%%%%%%%%%%%%%%%%%%%%%%%%%%%%%%%%%%%%%%%%%%%%%%%%%%%%%%%%%%%%%%
\setcounter{footnote}{0}
\renewcommand{\thefootnote}{\arabic{footnote}}
\renewcommand{\theequation}{\arabic{equation}}
%--------------------------------------------------------------------
%\section{Introduction}\label{sec1}
%\setcounter{section}{1}
\setcounter{equation}{0}
\indent
%-------------------------------------------------------------------
%%%%%%%%%%%%%%%%%%%%%%%%%%%%%%%%%%%%%%%%%%%%%%%%%%%%%%%%%%%%%%%%%%%%
%                                                                  %
%                              Text                                %
\indent                                                            %
%%%%%%%%%%%%%%%%%%%%%%%%%%%%%%%%%%%%%%%%%%%%%%%%%%%%%%%%%%%%%%%%%%%%
 
Nearly a decade ago, a series of $q$-analogues of the Virasoro algebra 
were investigated through analyzing an infinite set of $q$-deformed 
differential operators~\cite{saito,KS}. Some of these algebras can be 
organized as a $N=1$ supersymmetric algebra~\cite{CP1,BC}, and there 
exists a $N=2$ extension as well~\cite{N2}. As to the nonsupersymmetric 
parts, the following things are known. These types of $q$-deformed 
Virasoro algebras seem to act as $W$-infinity algebras on the space of 
soliton solutions~\cite{KP}, and their decompositions into the FFZ 
algebra~\cite{FFZ} are certainly possible at the level of the 
differential operator realizations. However, these suggested 
equivalences are not obvious so far in various observations at the 
level of field realizations: Sugawara constructions in terms of 
$q$-oscillators~\cite{CP1,CP2}, OPE representations~\cite{OPE1,OPE2}, 
and central extensions~\cite{CP1,OPE2}. In addition, none of 
realization-independent map relations is known yet, and it is important 
to examine relations between various realizations. An interesting 
remark is that one of these deformed algebras~\cite{saito}-\cite{CP1} 
is certainly a special case of the other (quantum) deformed 
Virasoro algebra emerged from the context of a lattice model~\cite{qkz}.

In this paper, we study the deformed super Virasoro algebra 
(Chaichian-Pre\v{s}najder type)~\cite{CP1} from a bit different point 
of view. Apart from the above equivalence problem, it is also an 
interesting question whether or not a supersymmetric extension of the 
algebra can really match with the concept of a physical (magnetic) 
deformation as mentioned below. The super algebra~\cite{CP1} consists of 
the commutation relations (called the algebras $q$-$Vir^F$ and $q$-$Vir^B$ 
in \cite{BC}) and the other parts involving supergenerators. 
In \cite{Hall}, it is shown that the algebra $q$-$Vir^F$ emerges as a 
natural generalization of the quantum algebra ${\cal U}_q(sl(2))$ in 
an electron system subjected on a two-dimensional surface in a uniform 
magnetic field~\cite{kogan,sato}. In this system, rather than the 
usual translation, the translation accompanied by a gauge 
transformation factor (the magnetic translation~\cite{mag}) plays 
an important role.

A linear combination of the magnetic translations forms the 
algebra $q$-$Vir^F$ (with no central extension). This is contrast to 
the fact that translational invariance (energy-momentum tensor) is 
related to the Virasoro algebra. {}~Furthermore, it is an interesting 
framework that a magnetic lattice becomes continuous as a magnetic field 
vanishes and then the $q=1$ case (Virasoro algebra) recovers 
in this limit. It is curious to examine whether or not this similarity 
would hold in a supersymmetric case, and hence we construct a couple 
of realizations of the supersymmetric extension of $q$-$Vir^F$ 
and $q$-$Vir^B$ in terms of the magnetic translation operators. 
Note that we only deal with centerless algebras, since the magnetic 
translations are differential operators.  

The magnetic translations defined on a two-dimensional lattice $(k,n)$; 
$k,n\in {\bf Z}$, satisfy the relation 
\beq{.1}
 T_{(k,n)}  T_{(l,m)} = q^{ln-mk\over2}(q-q^{-1})^{-1} T_{(k+l,n+m)} \ ,
\eeq
with realizing the FFZ algebra~\cite{FFZ}
\beq{.2}
 [\,T_{(k,n)}\,,  T_{(l,m)}\,] = [{ln-mk\over2}]_q T_{(k+l,n+m)} \ ,
\eeq
where 
\beq{qx}
[x]_q=(q^x-q^{-x})/(q-q^{-1})\ . 
\eeq
These relations are also appeared in the recent studies of 
non-commutative field theory~\cite{NC}. Hereafter, for the 
generality of discussion, we assume that the $T_{(k,n)}$ are defined 
in an abstract sense. \\

{\underbar {\tt The algebra $q$-$Vir^F$}}\\

The algebra $q$-$Vir^F$ is defined by 
\beq{qvirF}
[\, F_n^{(k)}\,, F_m^{(l)}\,]={1\over2}\sum_{\eps,\eta=\pm1}
[{\eps nl -\eta mk\over2}]_q {[\eps k+\eta l]_q\over[k]_q[l]_q}
\, F^{(\eps k +\eta l)}_{n+m} \ ,
\eeq
which is the maximal symmetric form in the generator indices~\cite{KP}. 
The upper and lower indices on $F_n^{(k)}$ take all integers, however 
for later convenience, we may exclude $k=0$ without any contradiction. 
The central extension of this algebra can be realized by the Sugawara 
construction of fermionic oscillators~\cite{CP1}. If we assume the relation 
\beq{.Fk}
F_n^{(k)} = F_n^{(-k)}\ ,
\eeq
the above algebra reduces to the following form:
\beq{qvirF2}
[\, F_n^{(k)}\,, F_m^{(l)}\,]= \sum_{\eps =\pm1}
[{\eps nl - mk\over2}]_q {[ k+\eps l]_q\over[k]_q[\eps l]_q}
\, F^{(k +\eps l)}_{n+m} \ .
\eeq
If we consider the $q\ra1$ limit with assuming
\beq{limF}
{}~F^{(k)}_n \quad\ra\quad L_n\ ,
\eeq
the algebra $q$-$Vir^F$ becomes the Virasoro algebra
\beq{vir}
[\, L_n\,,L_m\,] = (n-m) L_{n+m}\ .
\eeq 

There are two magnetic translation operator realizations for 
the $q$-$Vir^F$ generators; one is ~\cite{Hall} 
\beq{F1}
F_n^{(k)} = {1\over[k]_q}\sum_{\eps=\pm1}\eps\, T_{(\eps k,n)}\ ,
\qquad (k\not= 0)
\eeq
and the other is merely given by interchanging the roles of the 
two components of lattice coordinates $(n,m)$; 
\beq{F2}
F_n^{(k)} = {-1\over[k]_q}\sum_{\eps=\pm1}\eps\, T_{(n,\eps k)}\ ,
\qquad (k\not= 0)\ .
\eeq
Identifying
\beq{.J}
J_m^{(l)} = T_{(l,m)} \quad\mbox{or}\quad T_{(m,l)}
\quad\qquad\mbox{for Eqs.}\eq{F1}\quad\mbox{or}\quad\eq{F2}\ ,
\eeq
the following relation is satisfied in each case:
\beq{FJ}
[\, F_n^{(k)}\,, J_m^{(l)}\,] = {1\over[k]_q}\sum_{\eps=\pm1}\eps\,
[{nl-\eps mk\over2}]_q J_{n+m}^{(\eps k+l)}\ .
\eeq
This represents an analogue of the commutation relation between 
$u(1)$ currents and the Virasoro generators
\beq{VJ}
[\, L_n\,, J_m\,]=-m J_{n+m} \ .
\eeq
Here we put a remark. In the realization of $q$-$Vir^F$ by ghost 
oscillators, there exists the following closed algebra~\cite{OPE2} 
(in addition to Eq.\eq{qvirF2}): 
\beq{FR}
[\, F_n^{(k)}\,, R_m^{(l)}\,]={1\over[k]_q[l]_q} \sum_{\eps =\pm1}
[{\eps nl - mk\over2}]_q  [ k+\eps l]_q 
\, R^{(k +\eps l)}_{n+m} \ ,
\eeq
\beq{RR}
[\, R_n^{(k)}\,, R_m^{(l)}\,]={1\over[k]_q[l]_q} \sum_{\eps =\pm1}
[{\eps nl - mk\over2}]_q [ k+\eps l]_q
\, F^{(k +\eps l)}_{n+m} \ .
\eeq
In the present case, we can realize the generators $R_n^{(k)}$ as 
\beq{R}
R_n^{(k)} = {1\over[k]_q}\sum_{\eps=\pm1}\, T_{(\eps k,n)}\ ,
\qquad (k\not= 0)\ .
\eeq

{\underbar {\tt The algebra $q$-$Vir^B$}}\\

The other counterpart (bosonic) algebra is the algebra 
$q$-$Vir^B$~\cite{KS,CP1}:
\beq{qvirB}
[\, B_n^{(k)}\,,B_m^{(l)}\,]={1\over2}\sum_{\eps,\eta=\pm1}
[{n(\eps l+1)-m(\eta k+1)\over2}]_q B_{n+m}^{(\eps k+\eta l +\eps\eta)}\ .
\eeq
In contrast to $q$-$Vir^F$, the central extension of this algebra can be 
realized by the Sugawara construction of bosonic oscillators~\cite{CP1}. 
Note that there are two ways of taking the $q\ra1$ limit:
\beq{limB1}
B^{(k)}_n \quad\ra\quad L_n\ ,
\eeq
\beq{limB2}
B^{(k)}_n \quad\ra\quad k\,L_n\ ,
\eeq
where both limits satisfy the Virasoro algebra \eq{vir}.

We here present the following four magnetic translation operator 
realizations (let them referred to as ${\cal R}_1^{\pm}$ 
and ${\cal R}_2^{\pm}$):
\beq{R1}
{\cal R}_1^{\pm}\,: \quad
B_n^{(k)} = {1\over2}\sum_{\eps,\eta=\pm1}\eps\, q^{\pm\eps n\over2}
T_{(\eta k+\eps,n)}\ ,
\eeq
\beq{R2}
{\cal R}_2^{\pm}\,: \quad
B_n^{(k)} = {1\over2}\sum_{\eps,\eta=\pm1}\eta\, q^{\pm\eps n\over2}
T_{(\eta k+\eps,n)}\ .
\eeq
The deformed $u(1)$ currents are identified for these realizations 
as follows: 
\beq{.j}
J_m^{(l)} = T_{(\pm l,m)} \qquad\mbox{for}\quad 
{\cal R}_a^\pm\qquad (a=1,2)  \ ,
\eeq
and then the commutation relations with $B_n^{(k)}$ for 
${\cal R}_1^\pm$ turn out to be 
\beq{bj1}
[\,B_n^{(k)},\, J_m^{(l)}\,]=
{1\over2}\sum_{\eps,\eta=\pm1}\,\eps\,q^{\eps n/2}\,
[{nl-m(\eta k+\eps)\over2}]_q\, J_{n+m}^{(\eta k+l+\eps)}\ ,
\eeq
and for ${\cal R}_2^\pm$,
\beq{bj2}
[\,B_n^{(k)},\, J_m^{(l)}\,]=
{1\over2}\sum_{\eps,\eta=\pm1}\,\eta\,q^{\eps n/2}\,
[{nl-m(\eta k+\eps)\over2}]_q\, J_{n+m}^{(\eta k+l+\eps)}\ .
\eeq
When we take the $q\ra1$ limits of these commutators, we have to 
assume \eq{limB1} for the realizations ${\cal R}_1^\pm$, and \eq{limB2} 
for the realizations ${\cal R}_2^\pm$, in order to properly reproduce 
the correct limit \eq{VJ}. This suggests that the realizations 
${\cal R}_1^\pm$ and ${\cal R}_2^\pm$ certainly possess a different 
meaning from each other, although both satisfy the same algebra 
$q$-$Vir^B$.\\ 

{\underbar {\tt Superalgebra }}\\

In addition to the commutators \eq{qvirF} and \eq{qvirB}, a 
supersymmetric generalization of those deformed algebras consists of 
the following (anti-) commutation relations~\cite{CP1,OPE2}:
\beq{FB}
[\,F_n^{(k)}, B_m^{(l)}\,]=0 \ ,
\eeq
\beq{FG}
[\,F_n^{(k)}, G_m^{(l)}\,]={1\over[k]_q(q-q^{-1})}\sum_{\eps=\pm1}
\eps\, q^{nl-\eps mk\over2}G_{n+m}^{(\eps k+l)} \ ,
\eeq
\beq{BG}
[\,B_n^{(k)}, G_m^{(l)}\,]={-1\over2(q-q^{-1})}\sum_{\eps,\eta=\pm1}
\eta\, q^{-n(l+\eta)+m(\eps k+\eta)\over2}G_{n+m}^{(\eps k+l+\eta)} \ ,
\eeq
\beq{GG}
\{\,G_n^{(k)}, G_m^{(l)}\,\}=2q^{(nl+mk)/2}B_{n+m}^{(k-l)}
+\sum_{\eps=\pm1}\eps q^{n(\eps-l)-m(k+\eps) \over2}
[k-l+\eps]_q F_{n+m}^{(k-l+\eps)}\ .
\eeq
This superalgebra was first proposed by Chaichian and 
Pre\v{s}najder~\cite{CP1}. The main issue of this paper is to realize 
this superalgebra in terms of the operators satisfying \eq{.1}. 
It is essential to introduce a fermionic freedom in order to 
express a superalgebra as usual. 
We thus use a pair of fermionic oscillators
\beq{osc}
\{\,b\,,\,b^{\dagger}\,\}=1\ , \qquad b^2=(b^{\dagger})^2=0\ . 
\eeq
{}~For example, these are realized by the Pauli matrices
\beq{mat}
b=\sigma_x + i\sigma_y=\pmatrix{0&0\cr
        1&0\cr},\quad
b^{\dagger}=\sigma_x - i\sigma_y=\pmatrix{0&1\cr
        0&0\cr},\quad
\eeq
in the Hamiltonian system of a charged particle confined on a 
two-dimensional surface:
\beq{.}
H={1\over2}(p-eA)^2+{1\over2}B\sigma_z\ ,
\qquad\sigma_{z}=\pmatrix{1&0\cr
                          0&-1\cr}.
\eeq
In the following, we only assume the relations \eq{osc} for the 
generality of the argument. 

Let us consider the realizations of supersymmetric versions 
of $F_n^{(k)}$ and $B_n^{(k)}$: 
\beq{Fhat}
 F_n^{(k)}= {\cal R}(F_n^{(k)}) \otimes b\, b^{\dagger} \ ,
\eeq
\beq{Bhat}
 B_n^{(k)} ={\cal R}(B_n^{(k)}) \otimes b^{\dagger}b \ ,
\eeq
where ${\cal R}$ stands for a certain realization in the case of the 
non-supersymmetric algebras. It is obvious that for a given realization 
${\cal R}$, Eqs.\eq{Fhat} and \eq{Bhat} satisfy the commutation 
relation~\eq{FB} as well as each of $q$-$Vir^F$ and $q$-$Vir^B$. 

The forms of $G_n^{(k)}$ depend on the choice of realization 
${\cal R}$. In this paper, we employ the realization \eq{F1} as 
${\cal R}$ for the $q$-$Vir^F$ part. {}~For the $q$-$Vir^B$ part, 
we have four candidates for ${\cal R}$, as shown in \eq{R1} and 
\eq{R2}. However we have found only two realizations, 
which satisfy the relations \eq{FG}, \eq{BG} and \eq{GG}. 
One is for the realization ${\cal R}_1^+$, 
\beq{G1}
G_n^{(k)}=\sqrt{q-q^{-1}}\Bigl(\,
\sum_{\eps=\pm1} \eps q^{\eps n/2}T_{(k+\eps,n)}\otimes b
+T_{(-k,n)}\otimes b^{\dagger}\,\Bigr)\ ,
\eeq
and the other is for the realization ${\cal R}_1^-$,
\beq{G2}
G_n^{(k)}=\sqrt{q-q^{-1}}\Bigl(\,
T_{(k,n)}\otimes b+
\sum_{\eps=\pm1} \eps q^{-\eps n/2}T_{(\eps-k,n)}\otimes b^{\dagger}
\,\Bigr)\ \ .
\eeq

In summary, we have presented the realizations of the deformed 
superalgebra given by \eq{qvirF}, \eq{qvirB} and \eq{FB}-\eq{GG}. 
The ${\cal R}({\cal F}_n^{(k)})$ is given by Eq.\eq{F1}, and 
${\cal R}({\cal B}_n^{(k)})$ is either ${\cal R}_1^+$ or ${\cal R}_1^-$ 
(see Eq.\eq{R1}), while $G_n^{(k)}$ are realized by Eqs.\eq{G1} 
or \eq{G2} respectively. {}~Finally, some remarks are in order.  \\

(i) The magnetic translation operator realizations lead only to the 
centerless algebras, whereas the normal orderings of $q$-deformed oscillators 
in the Sugawara construction lead to the central extensions~\cite{CP1}.\\
(ii) We have restricted ourselves to discuss the Ramond type generators, 
$G_n^{(k)}$; $n\in {\bf Z}$. However, the present results also apply 
to the Neveu-Schwarz type ($n\in {\bf Z}+1/2$), if one introduces 
another set of $T_{(n,k)}$ with half-integral indices 
like on a dual lattice. \\
(iii) The above four realizations of $q$-$Vir^B$ have been classified 
into two types; the realizations ${\cal R}^\pm_1$ satisfy the present 
superalgebra, while ${\cal R}^\pm_2$ do not. In addition, the former 
type realizes the commutation relation \eq{bj1}, which is different 
from \eq{bj2}. The role of the latter type should further be 
investigated. \\ 
(iv) The present superalgebra seems different from possible linear 
combinations of the super FFZ algebra, which does not assume the 
bilinear forms (such as $b\,b^\dagger$) for the nonsuperalgebra parts. 
The difference is clear if comparing with other simpler quantum 
superalgebra~\cite{sine}. \\
(v) If one wants to introduce a non-commutativity in the Grassmann 
space, the ordinary commutation relation~\eq{osc} should be replaced by 
deformed Grassmann operators like done in a previous work~\cite{gras}. 
However, this will probably lead to a different deformed algebra. 

%\vspace{5mm}
%\noindent
%{\bf Acknowledgement}
%\vspace{5mm}
%\noindent
%
%The author would like to thank 

%%%%%%%%%%%%%%%%%%%%%%%%%%%%%%%%%%%%%%%%%%%%%%%%%%%%%%%%%%%%%%%%%%%%%
%                      REFERENCES                                   %
%%%%%%%%%%%%%%%%%%%%%%%%%%%%%%%%%%%%%%%%%%%%%%%%%%%%%%%%%%%%%%%%%%%%%
%

\end{document}